\documentclass[journal] {IEEEtran}

\usepackage{bm}
\usepackage{array}
\usepackage[cmex10]{amsmath}
\usepackage{amsfonts}
\usepackage{verbatim}
\usepackage[dvipdfmx]{graphicx} 
\usepackage{epsfig}
\usepackage{amsthm}
\usepackage[caption=false,font=footnotesize]{subfig}
\usepackage{fixltx2e}
\usepackage{rotating}

\usepackage{url}
\usepackage{cite}

\usepackage[normalem]{ulem}
\usepackage{hyperref}
\usepackage{color,array,colortbl}
\definecolor{mygray}{gray}{.9}

\ifCLASSINFOpdf
\else
\fi

\begin{document}
%
\title{ A Functional Composition Approach to Filter Sharpening and Modular Filter Design}
%
%
%

\author{Sefa~Demirtas
        and~Alan~V.~Oppenheim
\thanks{Copyright (c) 2015 IEEE. Personal use of this material is permitted. However, permission to use this material for any other purposes must be obtained from the IEEE by sending a request to pubs-permissions@ieee.org.}%
\thanks{This paper has supplementary downloadable material available at \url{http://www.rle.mit.edu/dspg/pub_journal.html}, provided by the authors. This consists of the MATLAB code to generate Figures 1, 2 and 3 using the algorithms described in this paper and those in the related references. }%
\thanks{Sefa Demirtas was at the Massachusetts Institute of Technology at the time this work was completed. He is currently with Analog Devices Lyric Labs. e-mail: sefa@alum.mit.edu.}
\thanks{Alan V. Oppenheim is at the Massachusetts Institute of Technology. e-mail: avo@mit.edu.}
}

%
%

\markboth{Transactions on Signal Processing}%
{Demirtas and Oppenheim: A Functional Composition Approach to Filter Sharpening and Modular Filter Design}
%



\maketitle

\begin{abstract}
Designing and implementing systems as an interconnection of smaller subsystems is a common practice for modularity and standardization of components and design algorithms. Although not typically cast in this framework, many of these approaches can be viewed within the mathematical context of functional composition. This paper re-interprets and generalizes within the functional composition framework one such approach known as filter sharpening, i.e. interconnecting  filter modules which have significant approximation error in order to obtain improved filter characteristics. More specifically, filter sharpening is approached by determining the composing polynomial to minimize the infinity-norm of the  approximation error, utilizing the First Algorithm of Remez. This is applied both to sharpening for FIR, even-symmetric filters and for the more general case of subfilters that have complex-valued frequency responses including causal IIR filters and for continuous-time filters. Within the framework of functional composition, this paper also explores the use of functional decomposition to approximate a desired system as a composition of simpler functions based on a two-norm on the approximation error. Among the potential advantages of this decomposition is the ability for modular implementation  in which the inner component of the functional decomposition represents the subfilters and the outer the interconnection.
\end{abstract}

\begin{IEEEkeywords}
Functional composition and decomposition, modular filters, filter sharpening
\end{IEEEkeywords}

\IEEEpeerreviewmaketitle

\section{Introduction}

\IEEEPARstart{B}{uilding} large systems from an interconnection of smaller modules is a common practice in signal processing as this approach benefits from the relative simplicity of designing submodules, captures the capabilities and the sophistication of larger systems and often results in highly modular structures. One such application is filter sharpening \cite{Kaiser1977}, which corresponds to linear interconnections of replicas of a given subfilter to obtain improved overall frequency characteristics. The subfilters may, for example, be designed offline with desired precision and complexity. Filter sharpening provides a flexible alternative to designing a high-order sharp filter for a variety of specifications, for which each design would otherwise take valuable resources in the context of an application. Filter sharpening is currently utilized efficiently in a number of applications including prefilter and equalizer design \cite{Adams1984,Cabezas1989,Jiang1990} as well as more sophisticated decimation filters than those obtained by simple cascading \cite{Kwentus1997}. However, the traditional methods that have been proposed for filter sharpening \cite{Kaiser1977, Nakamura, Saramaki, Hartnett1995, Chen, Samadi2000} are rather restrictive in that they rely on and require the subfilters to be Type-I FIR filters\footnote{A Type-I FIR filter is a linear phase filter with the impulse response $h[n], n=0,1,2,\dots,2M$ that satisfies $h[n]=h[2M-n]$, and its frequency response can be expressed as a zero-phase response multiplied with $e^{-j\omega M}$ \cite{Oppenheim}. Such a filter can be time advanced by $M$ samples to obtain a non-causal filter with an even-symmetric and real-valued frequency response.} with real-valued coefficients, which are characterized in time domain by an even symmetry around an integer sample $M$ such that they have an even-symmetric and real-valued frequency response after a time shift by $M$. These methods consider a variety of optimality criteria such as yielding maximally flat responses around frequencies where the subfilter magnitude response is zero or unity \cite{Kaiser1977, Hartnett1995, Samadi2000} or minimizing the $\l_2$-norm of the approximation error \cite{Nakamura}. Even though a commonly preferred optimality criterion for filter approximations is the minimization of the maximum deviation from the ideal filter response, i.e. the $\l_{\infty}$-norm of the error, this has only been considered previously in \cite{Saramaki}.

In this paper, we revisit filter sharpening from a systematic point of view that re-interprets it in the framework of functional composition. This framework corresponds to the application of one function to the results of another function. Conversely, functional decomposition is directed at expressing a given function as a composition of other functions, usually of lower order or complexity. The approach in this paper, based on functional composition, removes the restrictions on the types of the filters that can be sharpened and also presents a systematic framework for designing modular filters with minimax optimality guarantees for this unrestricted set of subfilters. Furthermore, the functional composition framework utilizes a rich mathematical literature on polynomial decomposition that leads to methods for designing modular FIR filters without the need to specify a subfilter, at the expense of trading the minimax optimality guarantee for a locally optimal mean squared error solution.

Section \ref{sec:sharpening} reviews the traditional approaches to filter sharpening and discusses their shortcomings. Filter sharpening is expressed as functional composition in Section \ref{sec:composition} for which a set of methods are provided to obtain the optimal gains for the sharpening interconnection network. Section \ref{sec:decomposition} presents functional decomposition as a more general means than filter sharpening to obtain modular filters where a subfilter is not necessarily pre-specified. 

\section{Background}\label{sec:sharpening}
The traditional approaches to filter sharpening, beyond the strategy of cascading replicas, typically consider only Type-I FIR subfilters with real-valued coefficients, and embed their replicas within a network of adders and gains resulting in transfer functions with the form of a weighted sum of powers of the subfilter transfer function. More specifically, the zero-phase response of the resulting sharpened filter can be represented in the functional form $F(\tilde{G}(e^{j\omega}))$ where $\tilde{G}(e^{j\omega})$ is the even-symmetric zero-phase response of the subfilter to be sharpened, and $F(\cdot)$ is a polynomial reflecting the specifics of the interconnections. The traditional methods do not apply to a large class of IIR filters, continuous time filters or even to other types of FIR filters since they cannot be time shifted by an integer amount to obtain an even-symmetric zero-phase response.

Well known methods for sharpening a Type-I FIR filter with a zero-phase response $\tilde{G}(e^{j\omega})$ include cascading the filter with itself to obtain $\tilde{G}^2(e^{j\omega})$, and a more general approach based on twicing as proposed by  Tukey \cite{Tukey} which results in the effective zero-phase response $2\tilde{G}(e^{j\omega})-\tilde{G}^2(e^{j\omega})$. These methods reduce ripples in either the passband or the stopband while having the adverse behavior in the other band. Kaiser and Hamming \cite{Kaiser1977} refer to the polynomial $F(\cdot)$ as the amplitude change function and provide a general formula to yield higher order polynomials to sharpen $\tilde{G}(e^{j\omega})$ in both bands with a focus on yielding a maximally flat design around frequencies where $|\tilde{G}(e^{j\omega})|$ is zero or unity.

The results in \cite{Kaiser1977} on sharpening of Type-I FIR filters have led other authors \cite{Nakamura, Saramaki, Hartnett1995, Chen, Saramaki2006, Saramaki2009, Samadi2000} to approach this problem in a more structured way, often referring to the overall design after sharpening as a tapped cascaded interconnection of FIR subfilters. The method in \cite{Saramaki} constitutes an important benchmark to part of the work we present here when $\tilde{G}(e^{j\omega})$ is pre-specified since it can be interpreted in the form of functional composition. Furthermore, it considers the $\l_{\infty}$ norm for the approximation error to an ideal filter response and obtains the optimal sharpening coefficients. In order to illustrate the approach proposed in \cite{Saramaki}, consider a subfilter with a zero-phase response $\tilde{G}(e^{j\omega})$ satisfying
\begin{subequations}\label{eqn:framework::mapping}
\begin{align}
x_{p1}&\le \tilde{G}(e^{j\omega}) \le x_{p2}, \;\; \omega\in \Omega_P \label{eqn:framewok::subfilter_spec_pass}\\
x_{s1}&\le \tilde{G}(e^{j\omega}) \le x_{s2}, \;\; \omega\in \Omega_S \label{eqn:framewok::subfilter_spec_stop},
\end{align}
\end{subequations}
where $\Omega_P$ and $\Omega_S$ are the union of pass-band and stop-band frequency intervals, $x_{p1}$ and $x_{p2}$ are the minimum and maximum values of $\tilde{G}(e^{j\omega})$ in its passband, and $x_{s1}$ and $x_{s2}$ are the minimum and maximum values in its stopband, respectively. For a pre-specified order $K$ for the polynomial $F(x)$, sharpening with respect to the $\l_{\infty}$ norm reduces to finding the optimal $K^{th}$ order polynomial to approximate
\begin{equation}\label{eqn:SaramakiQ}
Q(x)=\left\{\begin{array}{lr}
1, & x_{p1} \le x \le x_{p2}\\
0,& x_{s1}\le x \le x_{s2}\\
\end{array} \right.
\end{equation}
with respect to the same norm. More specifically, for the composition $F(\tilde{G}(e^{j\omega}))$, $\tilde{G}(e^{j\omega})$ will map every value of $\omega$ in its passband to the interval $[x_{p1},x_{p2}]$ and $F(x)$ will map that value as close to unity as possible since it is the optimal $K^{th}$-order polynomial approximation to $Q(x)$. Therefore the composition $F(\tilde{G}(e^{j\omega}))$ approximates unity in the passband. The same argument follows for the stopband. This polynomial approximation problem can be solved directly using the Remez Exchange Algorithm. However, in \cite{Saramaki}, this has been recast as a Parks-McClellan FIR lowpass filter design problem by utilizing Chebyshev polynomials and introducing scaling and offset coefficients to $\tilde{G}(e^{j\omega})$ such that the inverse cosine of its extremum values correspond to the actual band edges of a prototype low pass filter, the solution of which in turn invokes the Remez Exchange Algorithm for a very efficient solution to determine the unique optimum.

Although the traditional methods proposed for sharpening filters have all emphasized the convenience of using several subfilters to build more sophisticated filters, they either involve restrictions on the subfilters or consider less preferable optimality criteria. For example, the method in \cite{Kaiser1977} uses its degrees of freedom to provide a flat response at frequencies where $\tilde{G}(e^{j\omega})$ is zero and unity. Although this method successfully suppresses sufficiently small ripples, it typically does not for larger ripples that are inherent in low order subfilters as it relies on vanishing higher order derivatives of the proposed amplitude change functions which can remain non-negligible in a Taylor series approximation in the vicinity of zero and unity. Moreover, the amplitude change function $F$ for a given order is fixed for any subfilter and is not customized based on the subfilter. The method in \cite{Nakamura} considers the $\l_2$-norm optimality, a criterion that is known to possibly lead to solutions with narrow but very large deviations from an ideal response. A commonly preferred norm for filter approximations, the $\l_{\infty}$-norm, that was considered in \cite{Saramaki} will also be the focus in this paper. Furthermore, all of these existing methods require a Type-I FIR subfilter and do not extend sharpening to more general filters such as non-symmetric filters, and discrete-time or continuous-time IIR filters, most of which do not have zero-phase responses after an appropriate time shift. The method developed in Section \ref{sec:composition} will be applicable to this most general case of subfilters.

\section{Functional Composition for Filter Sharpening}\label{sec:composition}
\subsection{Revisiting Sharpening as Composition}
In the functional composition form of $F(G(\cdot))$, $F$ and $G$ are unrestricted as long as the range set of $G$ lies in the domain on which $F$ is defined. As functional composition can efficiently capture and concisely represent a sequence of operations on an input, functional compositions are ubiquitous in several disciplines such as mathematics, computer science, and engineering and has been studied and exploited in different applications such as modeling deformable media in computer graphics \cite{Sederberg1986}, robotic arm manipulation \cite{Minimair2000, Gathen1995}, symbolic computation and root finding algorithms in mathematics \cite{Barton, Barton1985, Zippel1985}, creating artificial reverberations for audio \cite{Schroeder} and designing IIR filters as a tapped cascaded interconnection of identical allpass subfilters \cite{Saramaki1987} among many others.

Although it is natural and straightforward to interpret filter sharpening from a functional composition perspective, their analyses did not historically originate from this broader perspective and have not previously taken advantage of the underlying mathematics and structure. In this section, we introduce and explore an approach to filter sharpening which follows the formalism of functional composition, extending and generalizing our work in \cite{DemirtasAllerton2013}. This approach removes the restrictions on the types of filters that can be sharpened, considers minimax optimality guarantees, and provides an alternative and systematic perspective to the existing approaches. Moreover, the functional composition approach to sharpening allows the extension of the analysis to cases for which the composition is designed to approximate an ideal filter in its magnitude response rather than its total complex frequency response, a commonly used constraint, for example, when designing continuous-time filters.

Restricting $F$ to be a polynomial, the functional form of a sharpened filter transfer function will be a composition, $F(G(z))=\sum{f_kG^k(z)}$, where $G(z)$ is the transfer function of any subfilter and specifically is not restricted to being Type-I FIR or having a real-valued frequency response. The filter sharpening problem can be expressed as finding the optimal composing polynomial $F(\cdot)$ of a desired order $K$ that minimizes the error between the resulting filter response and the desired filter response. More specifically, in this section the coefficients $f_k$ of $F$ will be chosen to minimize the $\l_{\infty}$ norm of the approximation error,
\begin{equation}\label{eqn:freqresp::filter_sharp_optim}
\begin{aligned}
& \underset{\mathbf{f}}{\text{minimize}}
& & \Delta \\
& \text{subject to}
& &\left\|H(e^{j\omega})-\sum_{k=0}^K f_k G^k(e^{j\omega})\right\|_{\infty}\le \Delta,
\end{aligned}
\end{equation}
where $H(e^{j\omega})$ is the desired filter response. This problem can be solved in a straightforward manner for any finite set of frequency points using linear or convex optimization techniques. Methods for solving it on a continuum of frequency points on a closed (hence compact) subset of $\omega \in [-\pi,\pi]$ are discussed below. A very special subclass of this problem is that for which $G(e^{j\omega})=e^{-j\omega}$, in which case $F(G(e^{j\omega}))$ is the frequency response of the FIR filter $F(e^{-j\omega})$. This then corresponds to the traditional FIR filter design problem. The Parks-McClellan FIR filter design algorithm \cite{Parks1972} places a symmetry constraint on the coefficients of $F$ and solves this special case using the Remez Exchange Algorithm \cite{Cheney1966}. The coefficient symmetry constraint leads to the representation of the problem in terms of real sinusoids, which satisfy the Haar condition \cite{Cheney1966, Meinardus1967}, a restrictive condition required for the Remez Exchange Algorithm.

\subsection{Sharpening for a Desired Frequency Response}
We remove the coefficient symmetry and the Haar condition constraints and explore sharpening subfilters $G(e^{j\omega})$ that are more general than a unit delay by exploiting a less efficient but more general algorithm, namely the First Algorithm of Remez \cite{Cheney1966}, summarized in Algorithm 1. This algorithm solves the optimization problem
\begin{equation}\label{eqn:freqresp::general_optim}
\begin{aligned}
& \underset{\mathbf{f}}{\text{minimize}}
& & \Delta \\
& \text{subject to}
& &\left\|D(x)-\sum_{k=0}^K f_k U_k(x)\right\|_{\infty}\le \Delta,
\end{aligned}
\end{equation}
where $x$ takes values from a compact set $\mathcal{S}$, and $D(x)$ and $U_k(x),\,k=0,1,\dots,K$ are continuous functions on $\mathcal{S}$. More specifically, it yields the minimax-optimal linear combination coefficients $f_k$ for a set of continuous functions $U_k(x)$ to approximate a desired continuous function $D(x)$ on $\mathcal{S}$. A set of polynomial coefficients $f_k$ to optimally sharpen $G(e^{j\omega})$ can be obtained by regarding the parameter $x$ as the frequency $\omega$, setting $D(\cdot) = H(e^{j\omega})$, $U_k(\cdot) = G^k(e^{j\omega})$ and $\mathcal{S}$ as the union of passbands and stopbands which needs to be a closed subset of $[-\pi,\pi]$.

\begin{equation*}
\begin{aligned}
&\text{ALGORITHM 1: First Algorithm of Remez} \\
&\text{ \textbf{Input:} $U_k(x), k =0,1,\dots,K$; $D(x)$, and $\mathcal{S}$}\\
&\text{ \textbf{Output:} $\mathbf{f^{\ast}} = \arg\min_{\mathbf{f}}\left\|D(x)-\sum_{k=0}^K f_k U_k(x)\right\|_{\infty}$.}\\
&\text{Begin ($i=1$)}\\
\text{\bf{0.}}& \text{ Choose $\mathcal{S}^{[i]}=\{x_0,x_1,\dots,x_m\}\subset \mathcal{S}$ for any $m$ such that}\\
&\text{$m\ge K$ and the matrix  $[U_k(x_n)]_{k,n},\;k=0,1,\dots,K$;}\\
&\text{$n=0,1,\dots,m$ has column rank $K+1$.} \\
\text{\bf{1.}}& \text{ Set $\mathbf{f^{[i]}} = \arg\min_{\mathbf{f}} \left\{\max_{x \in \mathcal{S}^{(i)}} \left|D(x)-\sum_{k=0}^K f_k U_k(x)\right|\right\}$.}\\
\text{\bf{2.}}& \text{ Find $x^{[i]}=\arg\max_{x \in\mathcal{S}} \left|D(x)-\sum_{k=0}^K f^{(i)}_k U_k(x)\right|$.}\\
\text{\bf{3.}}& \text{ Set $\mathcal{S}^{[i+1]}\leftarrow \mathcal{S}^{[i]}\cup \{x^{[i]}\}$ and $i\leftarrow i+1$, go to Step 1.}\\
\end{aligned}
\end{equation*}

In Algorithm 1, the minimax error is guaranteed to converge to the optimal value and the algorithm yields a set of coefficients $f_k,\, k=0,1,\dots,K$ that attains this value even when $U_k(x), \, k=0,1,\dots,K$ do not satisfy the Haar conditions, or these function and $D(x)$ are not real-valued, unlike the requirements in the Remez Exchange Algorithm. In fact, when the Haar condition is not met, any clustering point of the sequence of parameter vectors $\mathbf{f}$ will attain the optimal solution. If the Haar condition is also satisfied, the iterative procedure will yield in the limit the unique optimal coefficients $f_k, \, k=0,1,\dots,K$ (\cite{Cheney1966}, page 97). Step 1 of the algorithm, since restricted to a finite and discrete set of points, is a linear optimization problem if the functions involved are real-valued, or a convex optimization problem if complex-valued, and can be easily solved. In this paper, we used the free packages CVX \cite{cvx,Grant2008} and YALMIP \cite{yalmip} for specifying and solving these optimization problems in a \texttt{MATLAB} environment. The algorithm can be terminated based on a pre-specified threshold on the change in the minimax error.

\begin{figure}
\centering
\includegraphics[scale =0.40] {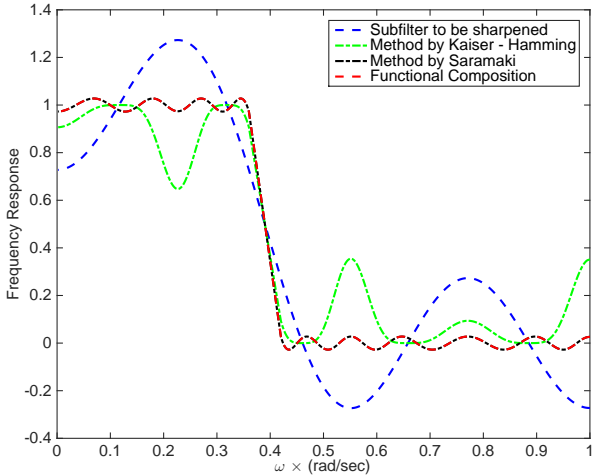}
\caption{The zero-phase response of a $10^{th}$-order Type-I FIR filter $G(e^{j\omega})$, and that of the resulting filters after sharpening with $7^{th}$-order polynomials. The polynomials were obtained using the method proposed in \cite{Kaiser1977}, the method proposed in \cite{Saramaki} and the functional composition approach stated in (\ref{eqn:freqresp::filter_sharp_optim}).}\label{fig:applications::comp_filt_sharp}
\end{figure}

Although functional composition for filter sharpening does not require $G(e^{j\omega})$ to be a Type-I FIR filter, the example in Figure \ref{fig:applications::comp_filt_sharp} is chosen with these constraints in order to show the improvement of the technique over the original filter sharpening method described in \cite{Kaiser1977}, which inherently carries these constraints. More specifically, in Figure \ref{fig:applications::comp_filt_sharp}, $G(e^{j\omega})$ is the zero-phase response of a $10^{th}$-order low-pass Type-I FIR filter obtained using the Parks-McClellan filter design algorithm with $\Omega_P=[0,0.36\pi]$ and $\Omega_S=[0.42\pi,\pi]$. This filter was sharpened using $7$-th order polynomials $F(\cdot)$ obtained using the method in \cite{Kaiser1977} and using the functional composition approach described here, which yields a superior sharpening particularly where the subfilter exhibits large ripples. In general, the relative improvement of the frequency response becomes more prominent as the subfilter exhibits larger deviations from the ideal response, which is usually the case with low order subfilters. A linear phase FIR filter with the same order as the $70^{th}$-order sharpened filters in the example of Figure \ref{fig:applications::comp_filt_sharp} can of course be designed directly with the Parks-McClellan algorithm, and would exhibit better frequency response characteristics than these sharpened filters. However, filter sharpening emphasizes building modular filters with relatively simple and low order subfilters that are straightforward to implement or readily available as opposed to designing a high order custom filter for each specific application.

Since the method in \cite{Saramaki} is also known to yield the minimax optimal sharpening error when $G(e^{j\omega})$ is the zero-phase response of a Type-I filter, the sharpened filter obtained using this method is also included in Figure \ref{fig:applications::comp_filt_sharp} as a benchmark. Although they utilize different tools, the functional composition approach and the method in \cite{Saramaki} both yielded the same approximating function even though the lack of the Haar condition for the set $\{G^k(e^{j\omega}), k=0,1,2,\dots,K\}$ suggests the optimal approximating function may not be unique. In order to gain further insight into this result, the representation of filter sharpening as functional composition can be used to show that the minimax-optimal $F$ is in fact unique and that both methods will yield the same sharpened filter for real-valued $G(e^{j\omega})$. More specifically, the filter sharpening problem given in (\ref{eqn:freqresp::filter_sharp_optim}) can be re-stated as

\begin{equation}
\underset{\mathbf{f}}{\mbox{minimize }} \underset{\omega\in \Omega_P\cup\Omega_S}{\mbox{max}}\left|H(e^{j\omega})-\sum_{k=0}^K f_k G^k(e^{j\omega})\right|
\end{equation}
which, if there exists a function $Q(x)$ that is continuous on $G(\Omega_P\cup\Omega_S)$ such that $Q\circ G(e^{j\omega})=H(e^{j\omega})$, is equivalent to
\begin{equation}
\underset{\mathbf{f}}{\mbox{minimize }} \underset{\omega\in \Omega_P\cup\Omega_S}{\mbox{max}}\left|\left(Q(x)-\sum_{k=0}^K f_k x^k\right)\circ G(e^{j\omega})\right|
\end{equation}
or
\begin{equation}
\underset{\mathbf{f}}{\mbox{minimize }} \underset{x\in G(\Omega_P\cup\Omega_S)}{\mbox{max}}\left|Q(x)-\sum_{k=0}^K f_k x^k\right|.
\end{equation}
Here, $G(\Omega_P\cup\Omega_S)$ denotes the image set of $G$ over the union of its passbands and the stopbands. For example, for the desired response $H(e^{j\omega})$ satisfying
\begin{equation}\label{eqn:idealH}
H(e^{j\omega})=\left\{\begin{array}{lr}
1, & \omega \in \Omega_P\\
0,&  \omega \in \Omega_S\\
\end{array} \right.,
\end{equation}
if $G(\Omega_P)$ and $G(\Omega_S)$ are disjoint sets as would be expected from any meaningful subfilter, $Q(x)$ in fact exists and becomes
\begin{equation}\label{eqn:myQ}
Q(x)=\left\{\begin{array}{lr}
1, & x \in G(\Omega_P)\\
0,&  x \in G(\Omega_S)\\
\end{array} \right.,
\end{equation}
which is equivalent to equation (\ref{eqn:SaramakiQ}) as obtained by the analysis given in \cite{Saramaki}. This manipulation implies that the filter sharpening problem reduces to the simple case of approximating $Q(x)$ with a $K$-th order polynomial $F(\cdot)$, which not only has a unique solution even if $\{G^k(e^{j\omega}), k=0,1,2,\dots,K\}$ does not satisfy the Haar conditions but also can be very efficiently solved using the Remez Exchange Algorithm as exploited in \cite{Saramaki}. Since the set of optimal sharpening coefficients is unique, both the functional composition method and the method in \cite{Saramaki} lead to the same solution for real-valued $G(e^{j\omega})$.

For complex-valued $G(e^{j\omega})$, solving for a polynomial approximation to $Q(x)$ in (\ref{eqn:myQ}) on $G(\Omega_P)$ and $G(\Omega_S)$ is not as straightforward as in the real case since these are subsets of the complex plane and not necessarily the real line. More specifically, the Remez Exchange Algorithm cannot be applied directly in this case. On the other hand, although not as efficient as the Remez Exchange Algorithm, the First Algorithm of Remez can still be used in this general case to find the optimal sharpening coefficients while none of the existing filter sharpening methods remain applicable except for simple cascading.

\subsection{Sharpening for a Desired Magnitude Response}
A general disadvantage of composing complex-valued functions when compared to those that are real-valued is the additional requirement of matching the phase of the approximating function $F(G(e^{j\omega}))=\sum_k f_kG^k(e^{j\omega})$ to that of $H(e^{j\omega})$. Due to this additional requirement, even the optimally-sharpened filter may not be satisfactory when the functional composition approach is applied directly. The approximation quality may improve significantly if only the magnitude response of the filter is desired to be approximated with that of a composition in an application. This relaxation of the phase matching constraint arises in certain signal processing contexts, for example in the design specifications of IIR filters both in discrete and continuous domains. These applications can potentially benefit from an extension of the functional composition approach to cases for which the approximation quality is specified with respect to the difference between $|H(e^{j\omega})|$ and $|F(G(e^{j\omega}))|$.

Consider a variant of the filter sharpening problem stated as
\begin{equation}\label{eqn:freqresp::filter_sharp_optim_mag}
\begin{aligned}
& \underset{\mathbf{f}}{\text{minimize}}
& & \Delta \\
& \text{subject to}
& &\left\|\left|H(e^{j\omega})\right|-\left|\sum_{k=0}^K f_k G^k(e^{j\omega})\right|\right\|_{\infty}\le \Delta,
\end{aligned}
\end{equation}
or equivalently as
 \begin{equation}\label{eqn:freqresp::general_optim_mag_phase_added}
\begin{aligned}
& \underset{\mathbf{f}}{\text{minimize}}
& & \Delta \\
& \text{subject to}
& &\left\| M(\omega)e^{j\Theta_{\mathbf{f}}(\omega)}-\sum_{k=0}^K f_k G^k(e^{j\omega})\right\|_{\infty}\le \Delta.
\end{aligned}
\end{equation}
with $M(\omega)=|H(e^{j\omega})|$ and $\Theta_{\mathbf{f}}$ is the phase function in the equality
\begin{equation}\label{eqn:freqresp::theta_def}
 \sum_{k=0}^K f_k G^k(e^{j\omega})=\left|\sum_{k=0}^K f_k G^k(e^{j\omega})\right|e^{j\Theta_{\mathbf{f}}(\omega)}.
\end{equation}
The problem stated in (\ref{eqn:freqresp::general_optim_mag_phase_added}) is no longer convex in $f_k,k=0,1,\dots,K$, and in its current form cannot be expressed and solved with the same approach used for the problem stated in (\ref{eqn:freqresp::filter_sharp_optim}). Algorithm 2 below provides an alternative iterative procedure for determining a locally optimal solution to this problem utilizing the approach for the problem stated in (\ref{eqn:freqresp::filter_sharp_optim}) in one of its steps. Similar to Algorithm 1, the iterations can be terminated based on a pre-specified threshold on the change in the approximation error.
\begin{equation*}
\begin{aligned}
&\text{\hspace{1in}ALGORITHM 2} \\
&\text{ \textbf{Input:} $G^k(e^{j\omega})$; $M(\omega)=\left|H(e^{j\omega})\right|$; an arbitrary $\Theta^{[0]}(\omega)$}\\
&\text{ \textbf{Output:} A local optimum for }\\
&\text{ $\mathbf{f^{\ast}} = \arg\min_{\mathbf{f}}\left\|M(\omega)e^{j\Theta_{\mathbf{f}}(\omega)}-\sum_{k=0}^K f_k G^k(e^{j\omega})\right\|_{\infty}$.}\\
&\text{ Set $i=1$.} \\
\text{\bf{1.}}& \text{ Set $\mathbf{f^{[i]}} = \arg\min_{\mathbf{f}}\left\|M(\omega)e^{j\Theta^{[i-1]}(\omega)}-\sum_{k=0}^K f_k G^k(e^{j\omega})\right\|_{\infty}$.}\\
\text{\bf{2.}}& \text{ Set $\Theta^{[i]}(\omega) = \arg\min_{\Theta(\cdot)}\left\|M(\omega)e^{j\Theta(\omega)}-\sum_{k=0}^K f_k^{[i]} G^k(e^{j\omega})\right\|_{\infty}$}\\
\text{\bf{3.}}& \text{ Set $i\leftarrow i+1$, go to Step 1.}\\
\end{aligned}
\end{equation*}

The first two steps of Algorithm 2 correspond to alternating projections of a function between the sets $\mathcal{U}$ and $\mathcal{V}$ where
\begin{equation}\label{eqn:freqresp::U}
\mathcal{U} = \{P(\omega) \mbox{ s.t. } P(\omega)=\sum_{k=0}^K a_k G^k(e^{j\omega}), a_k \in \mathbb{R} \}
\end{equation}
and
\begin{equation}\label{eqn:freqresp::M}
\mathcal{V} = \{R(\omega) \mbox{ s.t. } R(\omega)=M(\omega)e^{j\Theta(\omega)}, \mbox{ $\forall$ real }\Theta(\omega) \},
\end{equation}
which result in an iterative search for a function in the span of $\left\{G^k(e^{j\omega})\right\}$ that has a magnitude as close to $M(\omega)=|H(e^{j\omega})|$ as possible. The first step is equivalent to approximating a desired filter response $M(\omega)e^{j\Theta^{[i-1]}(\omega)}$ by sharpening $G(e^{j\omega})$, which can be formulated as (\ref{eqn:freqresp::filter_sharp_optim}) and solved using the First Algorithm of Remez as described previously. The optimal phase $\Theta(\omega)$ for the second step of Algorithm 2 can be shown to be the phase of the optimal approximating function $\sum_k f_k^{[i]} G^k(e^{j\omega})$ obtained in the first step. More specifically, for any $\omega$, the square of the objective function in the second step of Algorithm 2 becomes

\begin{IEEEeqnarray}{ll}\label{eqn:freqresp::projection_M}
&\left|M(\omega)e^{j\Theta(\omega)}-\sum_{k=0}^K f^{[i]}_kG^k(e^{j\omega})\right|^2\nonumber\\
=&\left|M(\omega)e^{j\Theta(\omega)}-\left|\sum_{k=0}^K f^{[i]}_kG^k(e^{j\omega})\right|e^{j\Theta_{\mathbf{f}}^{[i]}(\omega)}\right|^2\nonumber\\
=&\left|M(\omega)\right|^2+\left|\sum_{k=0}^K f^{[i]}_kG^k(e^{j\omega})\right|^2\nonumber\\
&-2\left|M(\omega)\right|\left|\sum_{k=0}^K f^{[i]}_kG^k(e^{j\omega})\right|\cos\left(\Theta(\omega)-\Theta_{\mathbf{f}}^{[i]}(\omega)\right)
\end{IEEEeqnarray}
where the first equality follows from the definition of $\Theta^{[i]}_{\mathbf{f}}(\omega)$ in equation (\ref{eqn:freqresp::theta_def}) and the second equality follows from the law of cosines. The same optimal choice of $\Theta(\omega)=\Theta_{\mathbf{f}}^{[i]}(\omega)$ minimizes this objective function for every frequency $\omega$, hence it is the solution for the second step of this algorithm.

It is well known that if the two sets $\mathcal{U}$ and $\mathcal{V}$ are both convex with a non-empty intersection, the sequence of functions obtained during this iterative procedure of alternating projections would converge to a function in $\mathcal{U}\cap \mathcal{V}$ yielding $\Delta=0$ in (\ref{eqn:freqresp::general_optim_mag_phase_added}), or, if the intersection is empty, converge to the closest point of $\mathcal{U}$ to $\mathcal{V}$ attaining the global minimum of $\Delta$. Although the lack of convexity in $\mathcal{V}$ prevents establishing such guarantees as in the First Algorithm of Remez, the minimax error at each iteration, denoted as $\Delta^{[i]}$, is a non-increasing sequence. Specifically, during the $i$-th iteration, the approximation error in Algorithm 2 satisfies \cite{Demirtas2014}

\begin{IEEEeqnarray}{lCl}\label{eqn:decreasing_error}
\Delta^{[i-1]}&=&\left\|M(\omega)e^{j\Theta^{[i-1]}(\omega)}-\sum_{k=0}^K f^{[i-1]}_kG^k(e^{j\omega})\right\|_{\infty}\nonumber\\
& \ge & \left\|M(\omega)e^{j\Theta^{[i-1]}(\omega)}-\sum_{k=0}^K f^{[i]}_kG^k(e^{j\omega})\right\|_{\infty}\nonumber\\
&\ge & \left\|M(\omega)e^{j\Theta^{[i]}(\omega)}-\sum_{k=0}^K f^{[i]}_kG^k(e^{j\omega})\right\|_{\infty}\nonumber\\
&= & \Delta^{[i]},
\end{IEEEeqnarray}
where the first inequality follows from the minimization at Step 1 and the second inequality follows from the minimization at Step 2. Since the sequence $\Delta^{[i]}$ is bounded below by zero, it is going to converge and possibly to a positive value $\Delta_{opt}$. Furthermore, choosing the same set of initial frequency points each time Step 1 invokes the First Algorithm of Remez guarantees to have the sequence of coefficient vectors  $\mathbf{f^{[i]}}$ to be bounded \cite{Demirtas2014}. 

\begin{figure}
\includegraphics[scale =0.40] {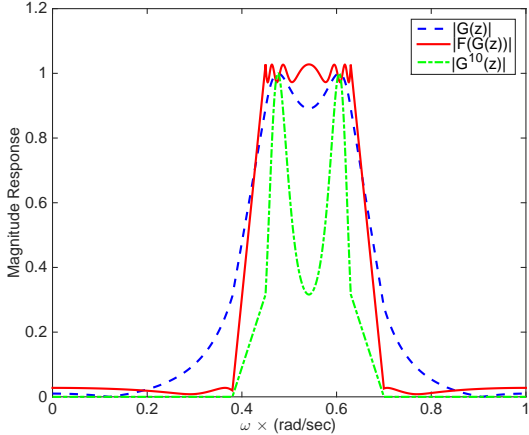}
\caption{The magnitude responses of a $4$-th order elliptic bandpass filter $G(z)$, the filter obtained by simply cascading $G(z)$ ten times and the modular filter obtained by composing $G(z)$ with a $10$-th order polynomial $F(\cdot)$ obtained by functional composition to approximate the desired response in its magnitude.}\label{fig:mag_filter_sharpening}
\end{figure}

Figure \ref{fig:mag_filter_sharpening} illustrates the magnitude response of a $4^{th}$-order elliptic bandpass subfilter $G(e^{j\omega})$ designed with passband edge frequencies of $0.45\pi$ and $0.63\pi$, maximum passband ripple of $1$dB and minimum stopband attenuation of $40$dB. Following the procedure in Algorithm 2, this filter is sharpened with a $10^{th}$-order polynomial $F(\cdot)$ to minimize $\left\||H(e^{j\omega})|-|F(G(e^{j\omega}))|\right\|_{\infty}$ rather than $\left\|H(e^{j\omega})-F(G(e^{j\omega}))\right\|_{\infty}$, where $|H|$ is unity on $\Omega_P=[0.45\pi,0.63\pi]$ and zero on $\Omega_S=[0,0.38\pi]\cup[0.70\pi,\pi]$. The response of the filter that is obtained by simply cascading ten replicas of the subfilter is also included in Figure \ref{fig:mag_filter_sharpening} for comparison. The functional composition approach to approximate the desired response in its magnitude yields a much better response when compared to the cascade.

\begin{figure}
\centering
\includegraphics[scale =0.40] {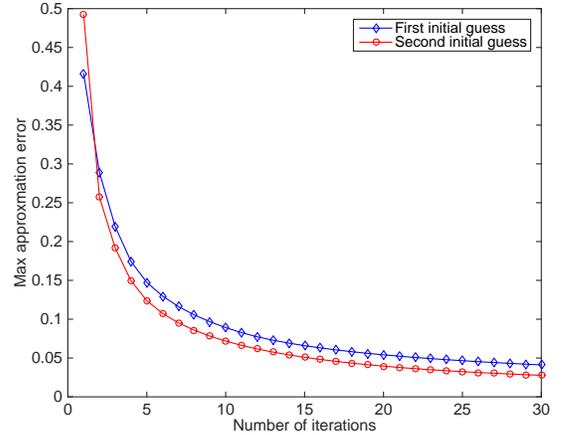}
\caption{The approximation error values $\Delta^{[i]}$ at each iteration in Algorithm 2 while sharpening the subfilter $G(z)$ given in Figure \ref{fig:mag_filter_sharpening} starting with two different initial phase functions $\Theta^{[0]}(\omega)$. Each of these were chosen as the phase of the function given by $\sum_{k=0}^{10}\tilde{f_k}G^k(e^{j\omega})$ with $\tilde{f_k}$ chosen randomly from a standard normal distribution.}\label{fig:applications::nonincreasing_error}
\end{figure}

Figure \ref{fig:applications::nonincreasing_error} illustrates the errors at each iteration of Algorithm 2 during the computation of the optimal coefficients for $f_k$ for the sharpening of $G(z)$ in the example of Figure \ref{fig:mag_filter_sharpening}. Starting at two different initial phase functions $\Theta^{[0]}(\omega)$, both curves have a non-increasing trend consistent with the analysis in (\ref{eqn:decreasing_error}). This figure also shows that different initial conditions lead to different initial errors as well as final error levels. Therefore, in such problems, different initial guesses may be tried until a satisfactory error level is achieved with increasing number of iterations. The coefficients for $F$ in Figure \ref{fig:mag_filter_sharpening} were chosen as those obtained by the procedure corresponding to the smaller error curve in Figure \ref{fig:applications::nonincreasing_error}.

An additional advantage of using composition for sharpening an IIR subfilter is stability. Composition with a polynomial $F(\cdot)$ introduces new zeros and no new poles, but only increases the multiplicity of the existing poles. This ensures that stability is not compromised through composition, a guarantee that lacks in practice in designing high order filters directly.

\section{Functional Composition and Decomposition for Modularity}\label{sec:decomposition}

\begin{figure}
\centering
\subfloat[]{\includegraphics[scale =0.60] {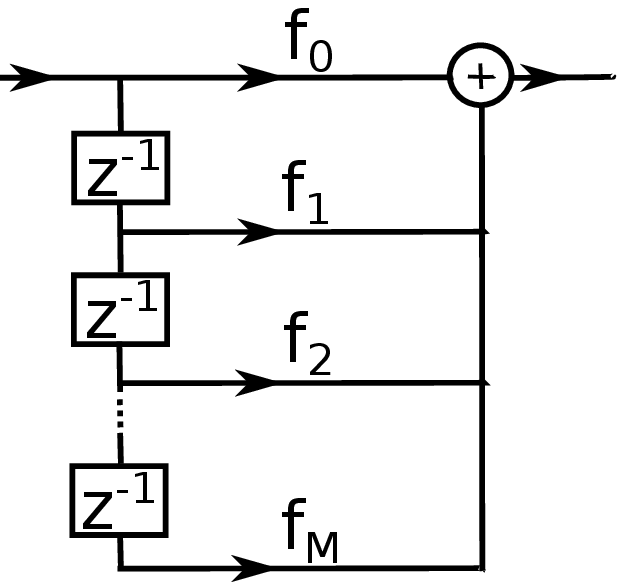}}
\quad
\subfloat[]{\includegraphics[scale =0.60] {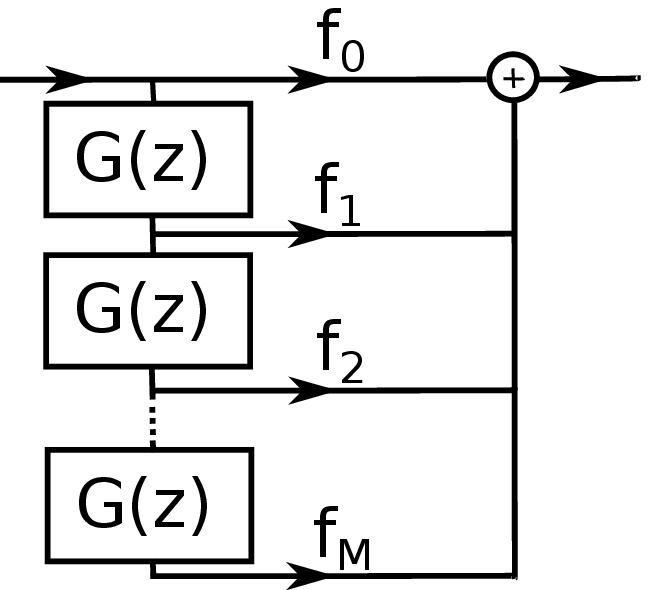}}
\caption{(a) The direct form implementation of an FIR filter $F(z)$ using a tapped delay line. (b) A generalized tapped delay line where the delays are replaced by another filter $G(z)$, often referred to as a tapped cascaded interconnection of subfilters.}\label{fig:applications::ComposedFIR}
\end{figure}

Sharpened filters obtained by functional composition as described in the previous sections can be implemented as a tapped cascaded interconnection of subfilters where each delay element in the direct form implementation of an FIR filter $F(z)$ is replaced by the subfilter $G(z)$ as illustrated in Figure \ref{fig:applications::ComposedFIR}. In addition to the motivating benefits of designing filters by sharpening simple subfilters, this structure has the advantage of being highly modular and flexible. For example, Nakamura \cite{Nakamura1985} proposed varying the tap coefficients and using different subfilters to obtain programmable FIR filters and adjustable magnitude responses. Moreover, the subfilters can be designed and fabricated offline with desired technology and accuracy. Although the multiplication rate increases in such structures, Saramaki \cite{Saramaki} emphasized the advantage of a reduced number of distinct multiplications and the possibility to use multiplexing in order to implement all subfilters using the same chip. From a design perspective, modular designs are also being increasingly promoted in VLSI designs where the overall system is often divided into either identical or few distinct sub-systems with a reduced emphasis on the number of multiplications or delay elements \cite{Mertzios1989}. This has the advantage of requiring a smaller number of different designs as well as the possibility of independent and efficient verification of sub-systems \cite{Mertzios1989, Vai2001}. 

In order to obtain modular filter structures when a subfilter is not pre-specified and hence functional composition cannot be invoked, functional decomposition techniques can be used to represent or approximate the desired response as a composition of simpler functions. A simple and suboptimal decomposition method is to approximate the desired response by a low order filter using a portion of the degrees of freedom that are available, with the remaining degrees of freedom used to sharpen this filter. Saramaki \cite{Saramaki} proposed a more systematic approach to obtaining a filter as a tapped cascaded interconnection of identical subfilters even when a subfilter is not pre-specified, which can be also viewed as composition. An alternative approach to designing modular filters without a pre-specified subfilter is to exploit known functional decomposition algorithms. A well studied class of functions for which several decomposition algorithms exist are polynomials \cite{Barton, Alagar,Kozen1989,Aubry2012, Corless, Giesbrecht, Botting2005}. These methods allow representing a given decomposable polynomial $H(z)$ as the composition of lower order polynomials $F(G(z))$, or approximating it with a decomposable one when it is non-decomposable. Modular filter design using polynomial decomposition techniques applied to FIR filters as examples follow. However this approach only yields locally optimum $\l_2$-error solutions for the cases in which an exact decomposition is unavailable. An overview of the most common approaches to exact and approximate decompositions of polynomials are given in \cite{Demirtas2013} and the sensitivities of composition and decomposition to coefficient perturbations are evaluated in \cite{Demirtas2012}.

\begin{figure}
\centering
\subfloat[]{\includegraphics[scale =0.35] {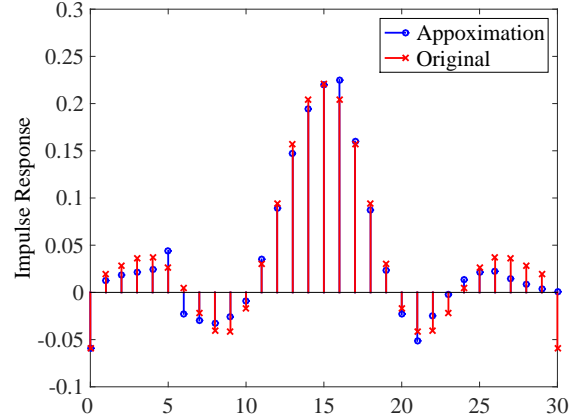}}\\
\subfloat[]{\includegraphics[scale =0.35] {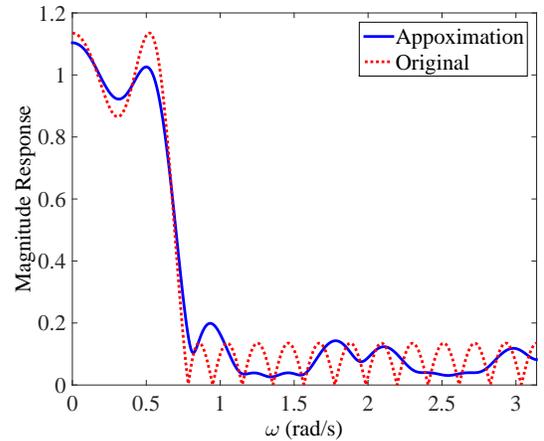}}
\caption{The comparison of a $30$-th order low-pass Parks-McClellan filter $H(z)$ with the passband and stopband edges of $0.20\pi$ and $0.24\pi$, respectively, with its approximate decomposition $F(G(z))$: (a) the impulse responses (b) the magnitude responses.}\label{fig:applications::approx_decomp_FIR}
\end{figure}

To illustrate  modular filter design using functional decomposition, consider a $30$-th order Parks-McClellan low-pass filter with the passband and stopband edges of $0.20\pi$ and $0.24\pi$, respectively. Figure \ref{fig:applications::approx_decomp_FIR}a shows the impulse responses of the original filter and its approximate decomposition\footnote{The functions in this section are polynomials in $z^{-1}$, and composing these polynomials refer to replacing $z^{-1}$ with other polynomials.} $F(G(z))$ obtained by the method described in \cite{Corless} where

\begin{IEEEeqnarray}{lCl}
F(z)&= & -0.0526+0.0649z^{-1}-0.0359z^{-2}-0.0021z^{-3}\nonumber\\
&&+0.1160z^{-4}-0.0226z^{-5}+0.0049z^{-6}
\end{IEEEeqnarray}

and

\begin{IEEEeqnarray}{lCl}
G(z)&=&-0.1037+0.1759z^{-1}+0.2667z^{-2}+0.3432z^{-3}\nonumber\\
&&+0.4321z^{-4}+0.7834z^{-5}.
\end{IEEEeqnarray}
Figure \ref{fig:applications::approx_decomp_FIR}b depicts the corresponding magnitude responses. Although the approximate polynomial decomposition optimizes the approximation with respect to the $\l_2$ norm and the impulse responses differ significantly, the magnitude response of the approximation still exhibits the general characteristics of the original low-pass filter magnitude response. However, this similarity does not always hold in general due to the difficulty of finding a nearby decomposable polynomial to any given non-decomposable polynomial. Moreover, the approximation in this case does not have the symmetry in the coefficients thereby losing the desirable linear phase property of the Parks-McClellan filter.

One approach to approximating Type-I FIR filters consists of first expressing the original frequency response as a polynomial in $\cos\omega$ and then decomposing this polynomial rather than decomposing the $z$-transform directly. More specifically, the Fourier transform of an even-symmetric filter $H(e^{j\omega})$ with order $2L$ and symmetric around $n=L$ can be represented as
\begin{IEEEeqnarray}{lCl}
H_{shifted}(e^{j \omega})&{=}&\displaystyle \sum_{n=-L}^{L} h_{shifted}[n]e^{-j \omega}\nonumber\\
&&{=}h_{shifted}[0]{+}{\displaystyle \sum_{n=1}^{L}} 2h_{shifted}[n] {\cos n \omega}
\end{IEEEeqnarray}
after a time shift of $L$ samples, where the time shift can be reversed by appropriate buffering once the filter is designed. Expanding each term in the sum using the Chebyshev polynomials leads to a polynomial in $\cos\omega$ as in 
\begin{equation}
H_{shifted}(e^{j \omega})=\displaystyle \sum_{n=0}^{L} b_n (\cos \omega)^n.
\end{equation}
In other words, the frequency response of the time shifted filter becomes $B(\cos\omega)$, where $B$ is a polynomial with coefficients $b_n$ and of order $L$. An approximate decomposition obtained using any approximate polynomial decomposition method such as the ones in \cite{Giesbrecht, Corless, Demirtas2013} and given by
\begin{equation}
B(x)\approx \hat{B}(x)= F(G(x))
\end{equation}
suggests a modular representation of the FIR filter as a tapped cascaded interconnection of subfilters where coefficients of $F$ are the tap coefficients and $G(\cos\omega)$ corresponds to an even-symmetric subfilter. However, the frequency responses $B(\cos\omega)$ and $\hat{B}(\cos\omega)$ were significantly different in simulations even in cases where the coefficients of $\hat{B}(x)$ were a good approximation to those of $B(x)$. This is expected since, in general, the proximity of the coefficients of two polynomials with respect to the $\l_2$ norm implies that their values are close with respect to the same norm when evaluated on the unit circle due to Parseval's theorem, and not necessarily on the interval $[-1,1]\subset \mathbb{R}$ from which $\cos\omega$ assumes values.

\begin{figure}
\centering
\includegraphics[scale =0.35] {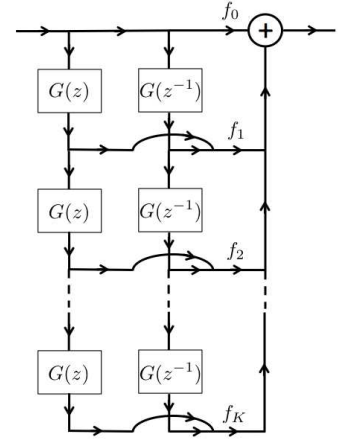}
\caption{The implementation of an even-symmetric FIR filter using an approximate decomposition of the form given in equation (\ref{eqn:applications::symmetric_decomp}).}\label{fig:applications::symmetric_decomp}
\end{figure}

In cases where the symmetry of a given filter is required to be preserved by the approximate decomposition, a third approach to performing the decomposition that also yields an acceptable approximation to the frequency response is to divide the impulse response before the decomposition into two subsequences which are related to each other through time reversal. More specifically, the $z$-transform of the time shifted filter can be expressed as
\begin{equation}
H_{shifted}(z) = C(z)+C(z^{-1}),
\end{equation}
where coefficients of $C(z)$ are those of $h_{shifted}[n]$ for $n\ge 0$ with the exception that its constant term is $\frac{h_{shifted}[0]}{2}$. An approximate decomposition of $C(z)$ as in
\begin{equation}
C(z)\approx F(G(z))
\end{equation}
yields
\begin{equation}\label{eqn:applications::symmetric_decomp}
H_{shifted}(z) \approx F(G(z))+F(G(z^{-1}))
\end{equation}
the coefficients of which are guaranteed to be symmetric. The implementation of this decomposable approximation leads to the modular structure given in Figure \ref{fig:applications::symmetric_decomp}. Although this implementation requires two different subfilters, namely $G(z)$ and $G(z^{-1})$, they are related through a time reversal which does not require the design of an additional subfilter. For on-line applications, this design can be used by introducing a buffer stage at the input to re-introduce causality.

As an illustration, the method of symmetric decomposition in equation (\ref{eqn:applications::symmetric_decomp}) was applied to the Parks-McClellan filter given in Figure \ref{fig:applications::approx_decomp_FIR}. The polynomial $C(z)$ corresponding to this polynomial is given by

\begin{IEEEeqnarray}{lCl}
C(z)&=&0.1105+0.2039z^{-1}+0.1572z^{-2}+0.0939z^{-3}\nonumber\\
&&+0.0307z^{-4}-0.0173z^{-5}-0.0412z^{-6}\nonumber\\
&&-0.0402z^{-7}-0.0215z^{-8}+0.0042z^{-9}\nonumber\\
&&+0.0260z^{-10}+0.0370z^{-11}+0.0364z^{-12}\nonumber\\
&&+0.0281z^{-13}+0.0192z^{-14}-0.0597z^{-15},
\end{IEEEeqnarray}
which was approximated as the composition of

\begin{IEEEeqnarray}{lCl}
F(z) &= &0.1862+0.2261z^{-1}+0.0020z^{-2}-0.0068z^{-3}\nonumber\\
&&-0.0132z^{-4}+0.0097z^{-5}
\end{IEEEeqnarray}
and
\begin{equation}
G(z)=-0.3359+0.8847z^{-1}+0.7099z^{-2}+0.4192z^{-3}.
\end{equation}

\begin{figure}
\centering
\subfloat[]{\includegraphics[scale =0.35] {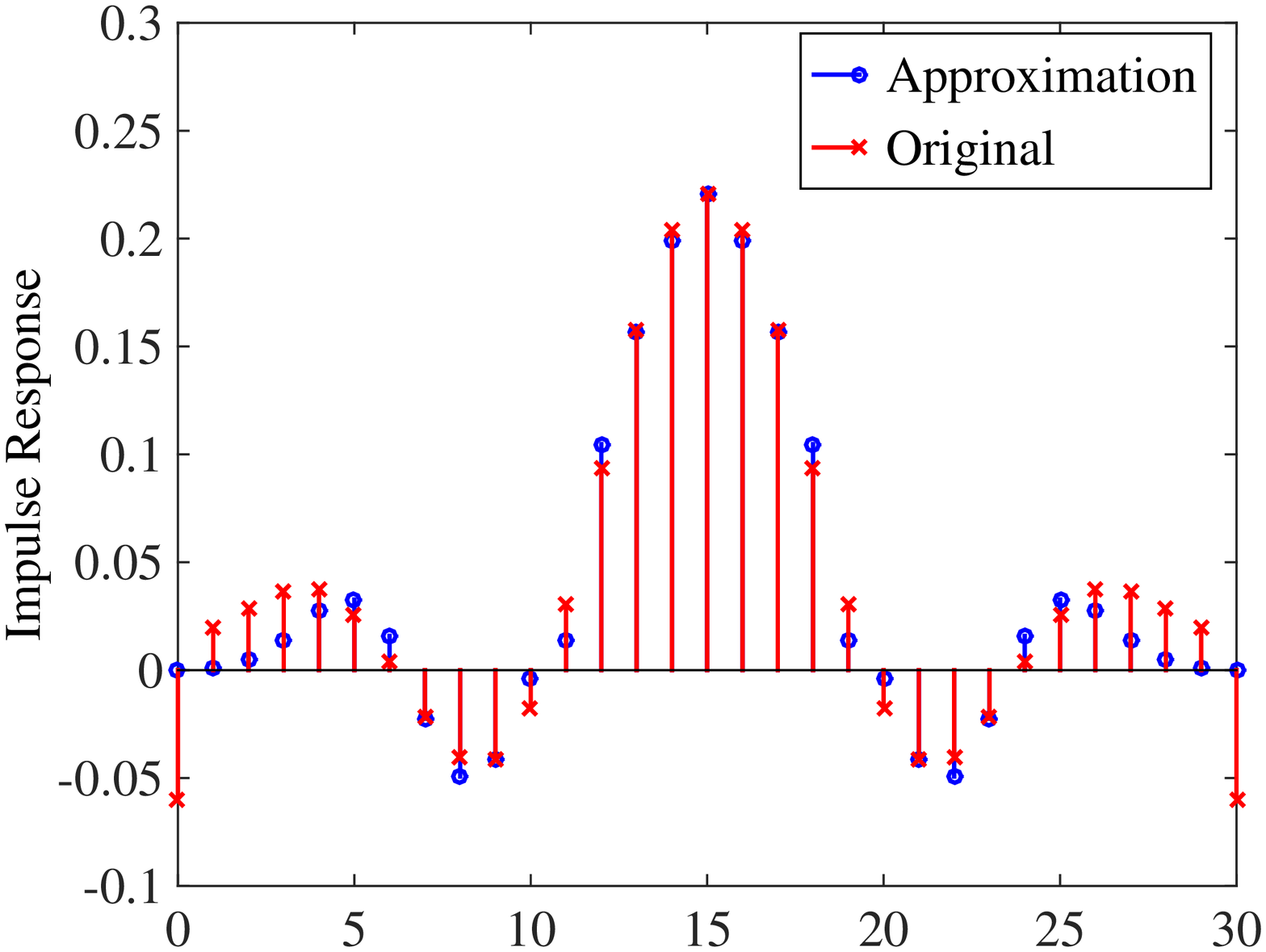}}\\
\subfloat[]{\includegraphics[scale =0.35] {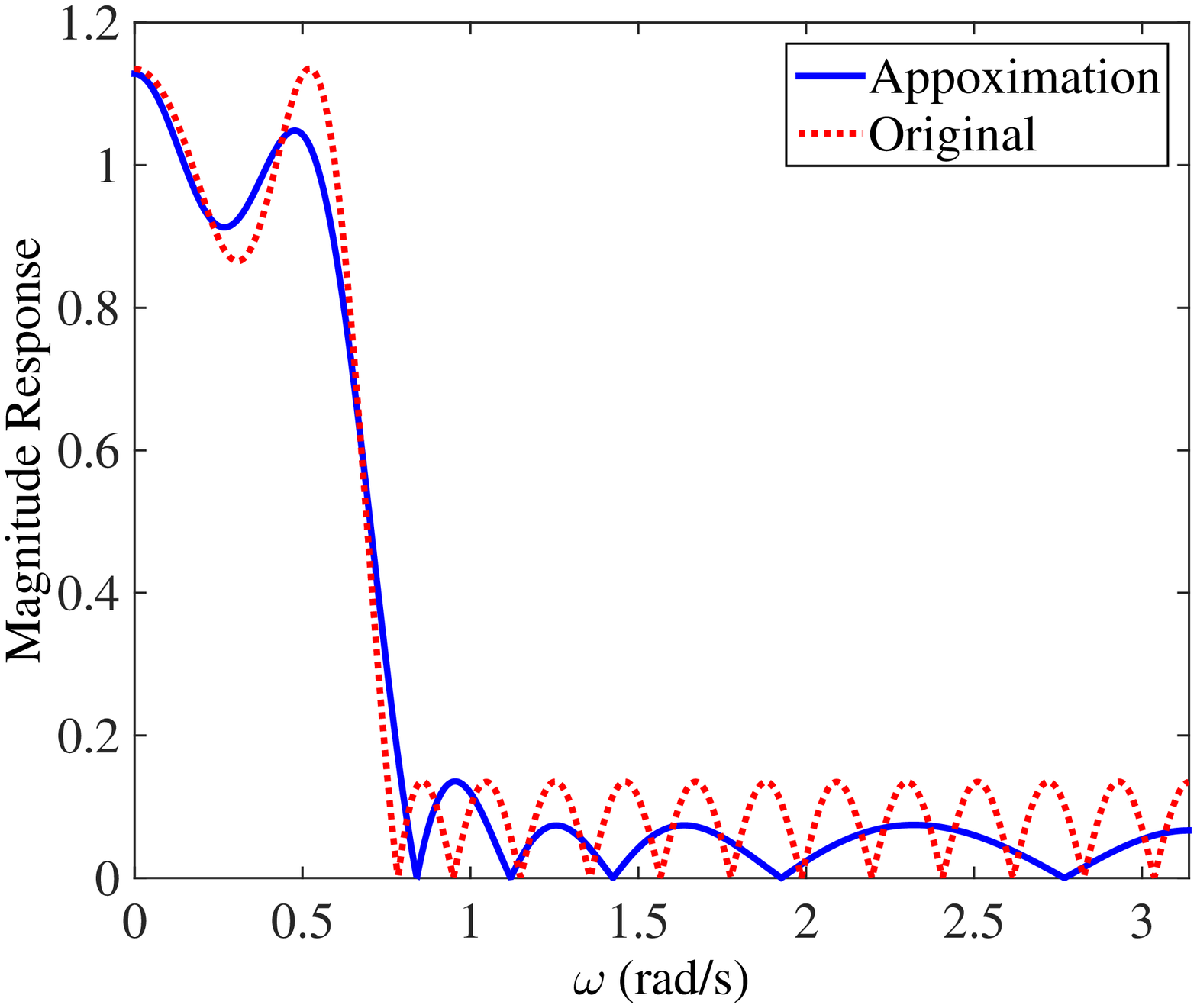}}
\caption{The comparison of the $30$-th order low-pass Parks-McClellan filter $H(z)$ with the passband and stopband edges of $0.20\pi$ and $0.24\pi$, respectively, with its approximate decomposition $F(G(z))+F(G(z^{-1}))$: (a) the impulse responses (b) the magnitude responses.}\label{fig:applications::symmetric_approx_decomp_FIR}
\end{figure}

Figure \ref{fig:applications::symmetric_approx_decomp_FIR}a illustrates the original response and its symmetric approximation obtained using this approach where the symmetry around $n=15$ was preserved as desired. As seen in Figure \ref{fig:applications::symmetric_approx_decomp_FIR}b which depicts the corresponding magnitude responses, the low-pass characteristics of the original filter were also preserved in this example with a slight widening of the transition region.

\section{Conclusions and Future Work}\label{sec:conclusions}
In this paper, functional composition is introduced as a broader and more systematic perspective in which to view, analyze and design modular filters and as an alternative to the traditional filter sharpening techniques. Algorithms for obtaining optimal gains to sharpen pre-specified filters are given. These algorithms can accommodate constraints on either frequency or magnitude responses. The functional composition view point removes the constraints on the subfilters as well as providing minimax ($\l_{\infty}$) optimality guarantees. Furthermore, functional decomposition is shown to be a broader approach and an alternative to obtaining modular filters even when a subfilter is not pre-specified.

Although not the focus in this paper, the same tools developed for filter sharpening can be applied to the more fundamental problem of designing FIR filters with nonlinear phase to minimize the maximum deviation from a desired response by selecting the subfilter as a unit delay, i.e. $G(e^{j\omega})=e^{-j\omega}$. The desired response $H(e^{j\omega})$ is also not restricted to being real-valued or piecewise-constant. Moreover, continuous-time filters for which the frequency response is defined on the entire real line can also be sharpened using these tools by transforming the problem into a compact frequency interval using the bilinear transformation. More specifically, the responses of both the subfilter and the desired filter defined on the real line can be mapped to the interval $[-\pi,\pi]$, an operation that preserves the minimax approximation error profile before invoking the First Algorithm of Remez.



\section*{Acknowledgment}

The authors would like to thank Guolong Su for implementing the approximate polynomial decomposition techniques to generate the numerical results in Section \ref{sec:decomposition}, and several anonymous reviewers for their helpful comments and suggestions. We also express our thanks to the Texas Instruments Leadership University Program, to Bose Corporation and to Analog Devices Inc. for their financial support.

\ifCLASSOPTIONcaptionsoff
  \newpage
\fi





\bibliographystyle{IEEEtran}
\bibliography{IEEEabrv,/Users/sdemirta/Documents/Dropbox/Personal/new_research/my_references}{}
%
%

%

\begin{IEEEbiography}{Sefa Demirtas}
Sefa Demirtas received the B.Sc. degree in electrical and electronics engineering (summa cum laude) from Bogazici University in Istanbul, Turkey, in 2007 and the M.Sc. and Ph.D. degrees in electrical engineering and computer science from the Massachusetts Institute of Technology (MIT), Cambridge, MA, in 2009 and 2014, respectively. While at MIT, he was a Research Assistant with the Digital Signal Processing Group (DSPG) and he held several Teaching Assistant positions in the EECS department mainly in graduate and undergraduate level signal processing classes.

Dr. Demirtas is currently a Research Scientist at Analog Devices Lyric Laboratories in Cambridge, MA. His current research interests lie broadly in signal processing and statistical inference and learning.
\end{IEEEbiography}

\begin{IEEEbiography}{Alan V. Oppenheim}
Alan V. Oppenheim received the S.B. and S.M. degrees in 1961 and the Sc.D. degree in 1964, all in Electrical Engineering, from the Massachusetts Institute of Technology.  He is also the recipient of an honorary doctorate from Tel Aviv University.  

In 1964, Dr. Oppenheim joined the faculty at MIT and is currently Professor of Electrical Engineering and Computer Science and a MacVicar Fellow His research interests are in the general area of signal processing, theory, algorithms and its applications.  He is coauthor of the widely used textbooks Discrete-Time Signal Processing (now in its third edition), Signals and Systems, and Digital Signal Processing, and the recently published textbook, Signals, Systems, and Inference.  He is also the editor of several advanced books on signal processing. In Spring 2015, an online version of his Discrete-Time Signal Processing course (6.341) was launched on edX through MIT.

Dr. Oppenheim is a member of the National Academy of Engineering, a fellow of the IEEE, and a member of Sigma Xi and Eta Kappa Nu.  He has been a Guggenheim Fellow and a Sackler Fellow.  He has received a number of awards for outstanding research and teaching, including the IEEE Education Medal, the IEEE Jack S. Kilby Signal Processing Medal, the IEEE Centennial Medal and the IEEE Third Millennium Medal. From the IEEE Signal Processing Society he has been honored with the Education Award, the Society Award, the Technical Achievement Award and the Senior Award. He has also received a number of awards at MIT for excellence in teaching, including the Bose Award, the Everett Moore Baker Award, and several awards for outstanding advising and mentoring.
\end{IEEEbiography}





\end{document}